\documentclass{aa}
\usepackage{graphicx}
\usepackage{amsmath}
\usepackage{widetext}
\usepackage{txfonts}
\usepackage[draft]{hyperref}

\begin{document} 
   \title{Narrowing the allowed mass range of ultralight bosons with the S2 star}
   \author{
      Riccardo Della Monica \inst{1}\thanks{email: rdellamonica@usal.es}
      \and
      Ivan de Martino\inst{1}\thanks{email: ivan.demartino@usal.es}
   }
   \institute{Universidad de Salamanca, Departamento de Fisica Fundamental, P. de la Merced S/N, Salamanca, ES}
   \date{Received XXX; accepted YYY}

  \abstract
   {}
   {It is well known that N-body simulations of ultralight bosons show the formation of a solitonic dark matter core in the innermost part of the halo. The scale length of such a soliton depends on the inverse of the mass of the boson. On the other hand, the orbital motion of stars in the Galactic Center depends on the distribution of matter whether be it baryonic or dark, providing an excellent probe for the gravitational field of the region. In this Letter we propose the S-stars in the Galactic Center as a new observational tool, complementary to other astrophysical systems, to narrow down the range of allowed values for an ultralight dark matter candidate boson mass.}
   {We  built mock catalogs mirroring the forthcoming astrometric and spectroscopic observations of S2, and we used a MCMC analysis to predict the accuracy down to which the mass of an ultralight boson may be bounded, and we showed that, 
   once complementary constraints are considered,  this analysis will help to restrict the allowed range of the boson mass.}
   {Our analysis forecasts the bound on the mass of an ultralight boson to be $< 10^{-19}$ eV at the 95\% of confidence level.}
   {}


   \maketitle

\section{Introduction}
\label{sec:introduction}

Dark matter is undoubtedly one of the most intriguing mysteries of modern astrophysics. Research aimed at revealing its fundamental nature has not given the desired results \citep{Bertone2018,deMartino2020,Salucci2021}. The solution  to such a puzzling mystery seems to reside in physics  beyond the Standard Model of elementary particles. However, the most promising candidate continues to escape detection {\em de facto} opening to the "no-WIMPs" era.  In this panorama, the ultralight bosons whose existence is motivated by string theory \citep{Arvanitaki2010} are one of the most promising alternatives.

The  ultralight boson dark matter, also known as Fuzzy Dark Matter (FDM), suppresses the galaxy formation below the de Broglie scale $k\sim 1/\lambda$ (corresponding to a mass of  $\sim 10^8 M_\odot$ \citep{Kawai2022}) and easily explains the large dark cores of dwarf galaxies and ultra-diffuse galaxies as solitons on the same scale. These properties offer a viable solution to two of the most crucial problems of the standard cold dark matter paradigm, namely the cuspy/core and the missing satellite problems, that would otherwise require a still not fully known feedback from baryon physics or new physics beyond the Standard Model \citep{deMartino2020}. However, the value of the mass of the boson, $m_a$, is still not unanimously known, and statistical tensions have recently arisen in the estimation of $m_a$ from the kinematics of dwarf galaxies and from the Lyman-$\alpha$ forest.

On one hand, a boson mass of $\sim 10^{-22} $ eV is favored by  the Jeans analysis of dwarf spheroidal galaxies \citep{Chen2017} and of the ultra-diffuse galaxies \citep{Broadhurst2020,Pozo2021}, and explains  the central motion of bulge stars in the Milky Way \citep{DeMartino2020PDU}. On the other hand,  using observations of the Lyman-$\alpha$ forest (Ly-$\alpha$), the suppression of the  cosmic structure growth favors a boson mass larger than $2\times 10^{-20}$ eV \citep{Kobayashi2017, Rogers2021}. Recently, an analysis based on the kinematic data of a sample of ultra-diffuse galaxies provides the first estimation of the boson mass that reconciles the galactic and cosmological probes \citep{Hayashi2021}. Evidently, the picture is not yet complete and self-consistent. Nevertheless, there is not yet agreement between the astrophysical and cosmological evidence, and there is not yet detection of a signature of such an ultralight boson. To this purpose, complementary observations and tests are required. For instance, the detection of the characteristic scalar field oscillation using Pulsar Timing Array experiments may provide a smoking gun for such an ultralight boson \citep{DeMartino2017,Porayko2018}.

Alternatively, we have focused on studying the orbits of the S2 star around the supermassive black hole (SMBH) in the center of the Milky Way. The study of the orbit of this star has been going on undaunted for three decades \citep{Genzel2010}. In recent years it has led to the measurement of relativistic effects such as Doppler and gravitational redshift, and orbital precession \citep{gravity2018, Gravity2020, Gravity2021, Do2019}. All measures confirm the predictions of General Relativity (GR) at more than $\sim 7\sigma$ \citep{Gravity2021}. In addition, the recent image of the shadow of the SMBH processed by the Event Horizon Collaboration  confirms the existence of a black hole in the center of the Milky Way \citep{EHT2022a, EHT2022f}. Clearly, the existing margin to find departures from GR is extremely narrow, and all the tests done so far continue to prove favorable to the predictions of GR rather than to a change in the theory of underlying gravity \citep{deMartino2021, DellaMonica2022, DellaMonica2022b}. Still, one may wonder what information can be extracted about the distribution of dark matter at these scales. 

Ultralight bosons form a solitonic core in the innermost part of each virialized halo \citep{Schive2014}. Depending on the inverse mass of the bosons, the core may be compact enough to add a non-negligible  amount of dark matter mass around the SMBH. On one hand, increasing the boson mass would make the soliton more compact adding more mass in the innermost part of the Galaxy. On the other hand, increasing the boson mass would reduce the size of the soliton allowing the S2 star to go in and out of the solitonic core along its eccentric orbit. While a more complete approach would involve constructing a rotating black hole metric embedded in a dark matter distribution coupled to the black hole metric, S2 allows us to work in the weak field limit where these effects can be linearized and added together. In what follows we have considered the acceleration of a test particle around a Schwarzschild black hole at the first post-Newtonian (1PN) order with an additional contribution due to the distribution of dark matter. Under these assumptions, we then used a mock catalog mimicking the accuracy of the GRAVITY instrument to forecast the allowed boson mass ranges.

\section{Ultralight dark matter halo around a supermassive black hole}
\label{sec:fdm_bh}

The dynamics of dark matter composed of ultralight bosons (with mass $m_a \sim 10^{-23\div-17}$ eV) 
can be described in terms of a scalar field $\phi$ minimally coupled to the space-time metric $g_{\mu\nu}$ for which, on galactic scales, the ultralight bosons (or axion-like particles) self-interaction and its coupling to ordinary matter can be neglected \citep{Hui2017}. In this case, the scalar field action reads as
\begin{equation}
    \mathcal{S} = \int \frac{d^4x}{\hbar c^2}\sqrt{-g}\left[\frac{1}{2}g^{\mu\nu}\partial_\mu\phi\partial_\nu\phi-\frac{1}{2}\frac{m^2c^2}{\hbar^2}\phi
    ^2\right].
\end{equation}
A large collection of such particles, sharing the same quantum state, can be described as a classical scalar field. As such, it is possible to express $\phi$ in terms of a complex scalar $\psi$, corresponding to the coherent wave function associated with the ultralight bosons in the non-relativistic regime (in which they fall due to their inherently high density). On turn, the dynamics of $\psi$ is regulated by the Schr\"{o}dinger equation
\begin{equation}
    i\hbar\frac{\partial}{\partial t}\psi = -\frac{\hbar^2}{2m_a}\nabla^2\psi+m_aV\psi,\label{eq:sp_1}
\end{equation}
where $V(\vec{r},t)$ is the gravitational potential. For a self-gravitating system of ultralight bosons, the gravitational potential $V$ can be computed from the density field associated to the wave function ($\rho = m_a |\psi|^2$) via the Poisson equation
\begin{align}
    \nabla^2V &= 4\pi G\rho.
    \label{eq:sp_2}
\end{align}
Equations \eqref{eq:sp_1} and \eqref{eq:sp_2} form the so-called Schr\"{o}dinger-Poisson (SP) system of equations \citep{Schive2014}. It describes the dynamics of a self-gravitating halo of FDM supported by an additional internal quantum pressure arising from Heisenberg's uncertainty principle. Numerical simulations of FDM halos regulated by the SP equations have extensively studied the properties of the stable configurations of the system \citep{Schive2014,Schive2014b}, {\em e.g.} the generation of coherent standing waves of dark matter in the center of gravitationally-bound systems. These prominent dark matter solitonic cores are surrounded by wave interference patterns whose azimuthal average follows a Navarro-Frenk-White (NFW) profile, while the radial profile of the solitonic core is well approximated by \citep{Schive2014, Mocz2017}
\begin{equation}
    \rho_s(r) = \frac{\rho_0}{(1+Ar^2)^{8}},
    \label{eq:soliton}
\end{equation}
where $A$ is related to the core radius $r_c$ via $A = 9.1\times 10^{-2}/r_c^2$, and $\rho_0$ corresponds to the central density of the halo:
\begin{equation}
    \rho_0 = 1.9\left(\frac{m_a}{10^{-23}\;\textrm{eV}}\right)^{-2}\left(\frac{r_c}{\rm kpc}\right)^{-4} \frac{M_\odot}{\rm pc^3}\,.
\end{equation}

Thus, the solitonic profile depends on two parameters: the boson mass $m_a$ and the core radius $r_c$, corresponding to the radial coordinate at which the density has dropped to $\sim 50\%$ of the central value $\rho_0$. The core radius, in turn, can be related to the virial mass of the entire halo via a scaling relation \citep{Schive2014b}
\begin{equation}
    r_c = 1.6\left(\frac{m_a}{10^{-22}\textrm{ eV}}\right)^{-1}\left(\frac{M_{\rm halo}}{10^9\,M_\odot}\right)^{-1/3}\textrm{ kpc}.
    \label{eq:scaling_relation}
\end{equation}

In our analysis, we set the Milky Way halo mass to  $M_{\rm halo} = 1.08 \times 10^{12}M_\odot$ from the GAIA satellite  \citep{Cautun2020} and thus reduce the distribution in Eq. \ref{eq:soliton} to a uni-parametric family of profiles, only depending on the boson mass $m_a$, by making use of the scaling relation in Eq. \ref{eq:scaling_relation}. Since we are concerned with the motion of test particles around the Galactic Center SMBH happening on a scale of $\sim 10^3$ AU, we can safely assume that the only contribution of the dark matter distribution to the orbital dynamics of stars is given by the innermost region described by the solitonic profile. This contribution results in an additional acceleration term, $\vec{a}_{\rm DM}$,   provided by the dark matter mass enclosed in the orbit of a test particle. Since we are working in the weak field limit of GR, this acceleration term can be linearly added to the 1PN acceleration term experienced by the particle due to the gravitational field of a Schwarzschild black hole given by \citep{Will2008, Gravity2020}
\begin{equation}
    \vec{a}_{\bullet} = -\frac{GM_\bullet}{r^3}\vec{r}+\frac{GM_\bullet}{c^2r^2}\left[\left(4\frac{GM_\bullet}{r}-v^2\right)\frac{\vec{r}}{r}+4\dot{r}\vec{v}\right]\,,
    \label{eq:PPN_acc}
\end{equation}
where $M_\bullet$ is the mass of the central SMBH and $\vec{r}$ and $\vec{v}$ are the position and the velocity of the test particle respectively. The total acceleration is thus given by
\begin{equation}
    \vec{a} = \vec{a}_{\bullet}+\vec{a}_{\rm DM}.
    \label{eq:acceleration}
\end{equation}
where $\vec{a}_{\rm DM} = a_{\rm DM}^r\vec{r}/r$ is the radial component of the additional acceleration given in Eq. \eqref{eq:acc_dm}. Here, we have assumed that the mass centroid of the dark matter distribution coincides with the SMBH, hence the acceleration due to halo $\vec{a}_{\rm DM}$ is directed along the radial direction $\vec{r}/r$, parallel to the first term in Eq. \eqref{eq:PPN_acc}. Moreover, we assume that the dark matter density profile is unperturbed by the presence of the point mass $M_\bullet$. While the dark matter dynamics in the innermost regions of the halo can be modified by the presence of the SMBH resulting in a more peaked profile (smaller core radius, greater central density)  \citep{Davies2020}, this effect is negligible for a halo mass of $M_{\rm halo}\sim 10^{12}M_\odot$  and the density profile in Eq. \eqref{eq:soliton} is still valid. 

\section{Data for the S2 star}

The S2 star is a bright B-type star in the nuclear star cluster of the Milky Way that orbits the compact radio and X-ray source SgrA*. Thanks to its high near-infrared (NIR) K-band (2.2 $\mu$m) magnitude of $m_K\sim 14$, its orbital period of $T\sim 16$ yr, and its almost face on orbital inclination, it has been possible to accurately measure the orbital motion of S2 probing the gravitational field of the central SMBH \citep{Gillessen2009, Gillessen2017, Do2019}. The most advanced instrument employed for observing the S2 star is GRAVITY, operated by combining all four 8.2 m telescopes at VLT into one big interferometer \citep{Gravity2017}. GRAVITY provides exquisite astrometry for S2, whose nominal rms uncertainty in optimal conditions is expected to be as low as $\sim 10\,\mu$as. This instrument has been indeed used from $\sim 2016$ on, allowing to detect relativistic effects such as Doppler and gravitational redshift, and the orbital precession with unprecedented precision \citep{gravity2018, Gravity2020, Gravity2021}. Unfortunately, astrometric data for the S2 star derived by the Gravity Collaboration \citep{Gravity2020} from GRAVITY interferometric observations are not publicly available. Therefore, we rely on a mock catalog (spanning two entire orbital periods of the S2 star) that contains synthetic astrometric observations of S2, mirroring the accuracy and observational strategy of GRAVITY, and mock spectroscopic measurements of radial velocities as could be measured by the integral field spectrograph SINFONI at VLT \citep{Eisenhauer2003, Bonnet2004}. Further details on how the catalog is built can be found in \citet{DellaMonica2022} and in the Appendix \ref{app:mock}.

\section{Orbital model for S2}

\begin{figure*}
    \centering
    \includegraphics[width = \textwidth]{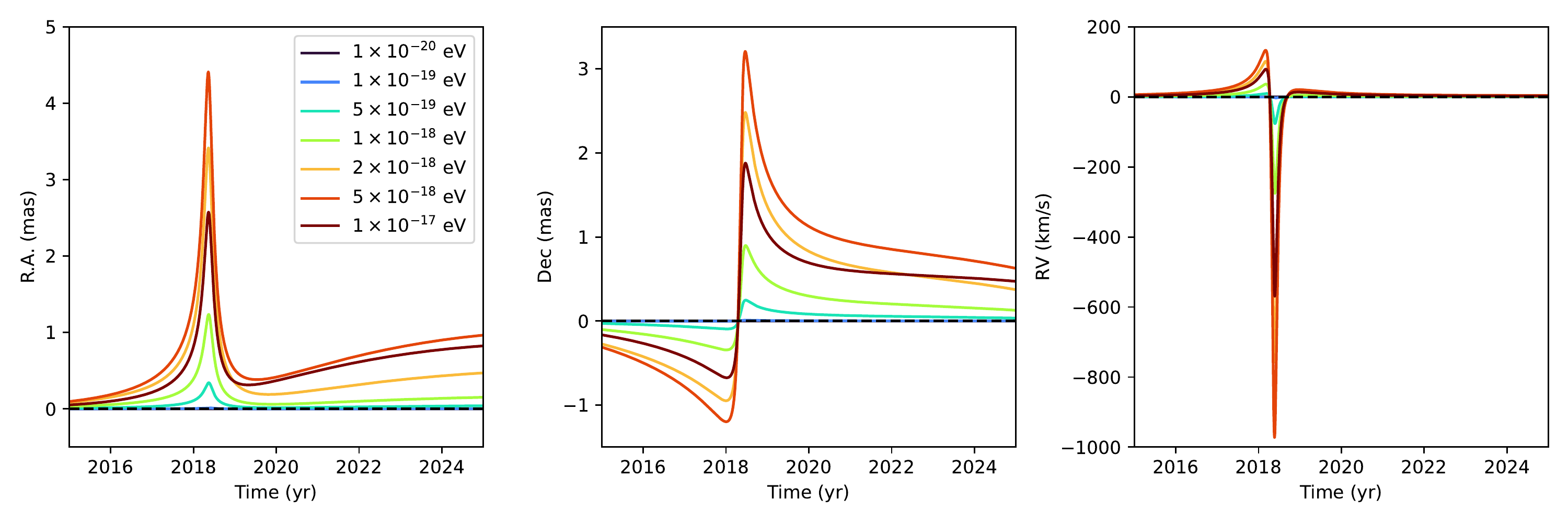}
    \caption{The deviation from a PPN orbit (no dark matter - dashed horizontal line) of the observable right ascension \emph{(left panel)}, declination \emph{(central panel)} and line of sight velocity \emph{(right panel)} of the S2 star for different values of the boson mass $m_a$.}
    \label{fig:s2_orbit_m_a}
\end{figure*}

A unique orbit for the S2 star can be determined by integrating the equations of motion given by the acceleration in Eq. \eqref{eq:acceleration}.
Initial conditions are assigned by means of the orbital elements of the osculating keplerian orbit at the initial time. In particular, the orbital period $T$, the time of pericenter passage $t_P$, the semi-major axis $a$, the eccentricity $e$, the inclination $i$, the longitude of the ascending node $\Omega$ and the argument of the pericenter $\omega$ altogether fix a unique value for the time of the apocenter passage $t_0 = t_P-T/2$ and for the initial state vector of the star ($x_0,y_0,z_0,{v}_{\rm x,0},{v}_{\rm y,0},{v}_{\rm z,0}$) at that time. 
Once integrated numerically, we need to convert the star's trajectory in the BH reference frame into observable quantities. In particular, the near-infrared (NIR) observations of the GC \citep{Gillessen2017, Gravity2020} provide the sky-projected astrometric positions of S2, while spectroscopic observations allow determining its line-of-sight velocity. Firstly, we perform a geometrical projection of the star trajectory into the observer reference frame by means of the Thiele-Innes elements \citep{Taff1986}. Then, by diving the physical positions of the star into this reference frame by the distance $D_\bullet$ of the observer from the GC, we compute the angular separations (relative right ascension and declination) between SgrA* and the star, as observed from Earth. Since the star is not always at the same distance from the observer (due to the inclination of its orbit) the time delay due to the different propagation time of the light emitted by the star, i.e. the Rømer delay, has to be taken into account (higher order effects of time delay, such as the Shapiro and Einstein time delay can be neglected for our purposes \citep{Do2019, DellaMonica2022}). For what concerns the spectroscopic observable, we need to take into account (\emph{i}) the kinematical line-of-sight velocity of the star which produces the frequency shift of the emitted light by the star; (\emph{ii}) the 3-dimensional kinematical velocity of the star that (especially at pericenter where $v\sim 7700$ km/s) causes a special relativistic time dilation and thus an additional redshift contribution; (\emph{iii}) the gravitational time dilation which occurs very close to the supermassive compact central object and reflects into an additional redshift component. Upon considering all the mentioned observational effects, our orbital model depends on 9 parameters: our distance ($D_\bullet$) from the GC; the 7 Keplerian elements ($T$, $t_P$, $a$, $e$, $i$, $\omega$, $\Omega$) and the boson mass $m_a$ (additional details on our orbital model are contained in Appendix \ref{app:initial_conditions}).

In order to assess the impact of the extended dark matter mass component, in Figure \ref{fig:s2_orbit_m_a} we report the astrometric and spectroscopic observable for the orbit of S2 around the last pericenter passage in $\sim 2018$, for different values of $m_a$, relative to the case without the dark matter contribution to the acceleration. In particular, we have fixed the values of the orbital parameters to the ones derived in \citet{Gravity2020} (in which only the 1PN acceleration in Eq. \eqref{eq:PPN_acc} is taken into account) and changed the value of $m_a$ over multiple orders of magnitude. As can be seen from the figure, for $m_a \lesssim 5\times^{-19}$ eV no deviation is observable on the orbit of S2. However, for greater values of the boson mass, the orbit starts to depart, with the greatest discrepancy from the case without dark matter at around $m_a \sim 5 \times 10^{-18}$ eV. This departure is detectable in all three observable quantities and its magnitude is related to both the density profile and the amount of enclosed mass within the orbit of S2. 

\section{Results}

We have explored the 9-dimensional parameter space of the orbital model by applying a Markov Chain Monte Carlo (MCMC) algorithm implemented in \texttt{emcee} \citep{emcee}. We adopted the uniform priors reported in Table \ref{tab:priors} that cover the range [$\mu-15\sigma$, $\mu+15\sigma$], where $\mu$ and $\sigma$ are the best fit values and uncertainties on each parameter resulting from the analysis in \citet{Gravity2020}. 
Finally, we have set a large flat prior on our parameter of interest, $m_a$, spanning over multiple orders of magnitude ($10^{-23}\textrm{ eV}\leq m_a \leq 10^{-17}\textrm{ eV}$) that we have sampled logarithmically in our analysis.

\begin{figure}
    \centering
    \includegraphics[width = 0.5\textwidth, keepaspectratio]{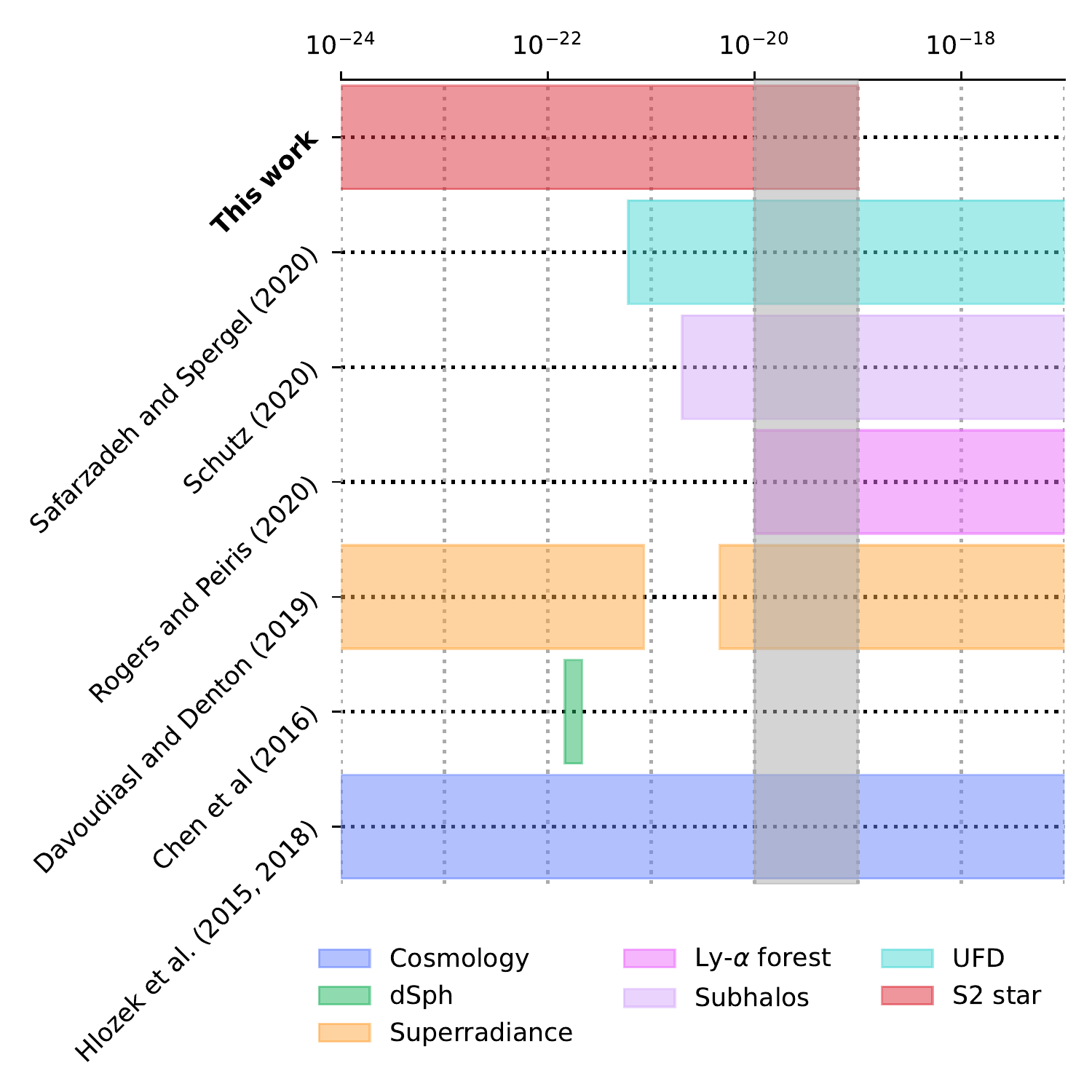}
    \caption{We compare the allowed range for the mass $m_a$ of ultra-light DM from our analysis with other constraints in literature coming from analysis at different scales. In particular: in blue we report constraints from cosmological observation (cosmic microwave background and large scale structures) from \citep{Hlozek2015, Hlozek2018}; in green, we report constraints from analysis in dwarf spheroidal galaxies (dSph) \citep{Chen2017}; purple boxes report constraints from the analysis of the Lyman-$\alpha$ forest from \citep{Rogers2021}; in orange, we report the allowed range for $m_a$ resulting from analysis of SMBH superradiance in \citep{Davoudiasl2019}; in aqua, the constraints from ultra-faint dwarf galaxies (UFD) from \citep{Safarzadeh2020} are reported and finally the lilac box reports the results of the analysis on subhalos from \citep{Schutz2020}. The vertical shaded strip corresponds to the intersection of the allowed intervals for $m_a$ from most of the astrophysical and cosmological probes with the exception of the ones coming from dSph.}
    \label{fig:mass_excluded}
\end{figure}

We report the  median value and the corresponding 68\% confidence intervals for all the parameters in Table \ref{tab:results_mock}. All the parameters in our orbital model will result by future observations of the S2 star to be bounded, except for $m_a$ on which it will only be possible to set an upper limit: $m_a \lesssim 1\times 10^{-19}$ eV at 95\% confidence interval.  Finally, Figure \ref{fig:mass_excluded} reports the allowed ranges of the boson mass from the literature. As soon as our prediction will be confirmed, one could be able to set an allowed range  of the boson mass, $10^{-20}\textrm{ eV}\le m_a \le 10^{-19}\textrm{ eV}$, that would agree with most of the astrophysical and cosmological probes except for the dwarf spheroidals as shown by the vertical shaded strip.

\begin{table}[]
        \setlength{\tabcolsep}{16.5pt}
        \renewcommand{\arraystretch}{1.5}
        \caption{The full set of median and 1$\sigma$ credible interval for the nine parameters of our model comes from our posterior analysis. For the parameter $m_a$ we report the 95\% confidence level upper limit.}
        \label{tab:results_mock}
        \begin{tabular}{|lcr|}
            \hline
            Parameter & Units & $1\sigma$ - credible interval \\ \hline
            $D_\bullet$ & (kpc) & $8.2454^{+0.0038}_{-0.0012}$\\
            $T$ & (yr) & $16.045502^{+0.000013}_{-0.000009}$\\
            $t_P - \textrm{J}2018.38$ & (yr) & $-0.000998^{+0.000012}_{-0.000009}$\\
            $a$ & (as) & $0.1250566^{+0.0000045}_{-0.0000019}$\\
            $e$ & () & $0.884649^{+0.0000051}_{-0.0000055}$\\
            $i$ & ($^\circ$) & $134.5678^{+0.001}_{-0.0025}$\\
            $\Omega$ & ($^\circ$) & $228.1707^{+0.003}_{-0.0027}$\\
            $\omega$ & ($^\circ$) & $66.2628^{+0.0025}_{-0.0023}$\\ \hline
            $m_a$ & (eV) & $\lesssim 1\times 10^{-19}$ {\footnotesize(95\% c.i.)}\\ \hline
        \end{tabular}
    \end{table}

\section{Discussion and conclusions}
Nowadays, the cold dark matter model is in a bind. On one hand, there are tensions between observations and numerical simulations that would require precise baryon feedback to be resolved \citep{deMartino2020}. On the other hand, the particle that serves as the best candidate continues to elude the efforts made for its revelation. Consequently, the possibility opens up that the dark matter paradigm has to be changed. And, in that regard, the ultralight bosons arising from a string theory landscape turn out to be an intriguing possibility \citep{Arvanitaki2010}. Such particles can resolve some of the long-standing tensions in the cold dark matter model \citep{deMartino2020}, and have a clear and unmistakable imprint that may soon be detected through future observations of pulsars at the Galactic center \citep{DeMartino2017,Porayko2018}. However, the mass of the boson favored by the kinematics of dwarf galaxies is in strong tension with that favored by Lyman-$\alpha$ \citep{Broadhurst2020,Pozo2021,Kobayashi2017}. In order to resolve this tension, a more accurate analysis of the kinematics of dwarf galaxies is necessary. In addition, complementary observations capable of probing the boson mass could provide key arguments for defining and narrowing the range allowed for the boson mass.

The orbital motion of stars at the Galactic center explores completely different astrophysical scales from those of dwarf galaxies or cosmological ones, providing in fact a complementary way to test the distribution of dark matter in the innermost part of the halo.  For instance, a boson mass of $\sim 10^{-18}$ eV would correspond to a solitonic core of $\sim 0.016$ pc that would be only about three times the semi-major axis of S2 with an enclosed mass of  $\sim0.8\times 10^3M_\odot$, which is about the $\sim0.2\%$ of the mass of the  SMBH. A higher boson mass would imply a smaller soliton, and therefore S2 would cross in and out of the solitonic core and its motion would result in a large departure from GR as shown in Figure \ref{fig:s2_orbit_m_a}. This is why this path is effective for the purpose of restricting the range allowed to the mass of the boson. And although we have shown that future observations will only allow us to set an upper limit on its mass, this will be enough to unprecedentedly narrow the existence range of ultralight bosons once complementary probes are taken into account.

\section*{Acknowledgments}
We thank the anonymous referee for his/her valuable comments that helped us to improve the presentation of our work.
RDM acknowledges support from Consejeria de Educación de la Junta de Castilla y León and from the Fondo Social Europeo. IDM acknowledges support from Ayuda  IJCI2018-036198-I  funded by  MCIN/AEI/  10.13039/501100011033  and:  FSE  “El FSE  invierte  en  tu  futuro”  o  “Financiado  por  la  Unión  Europea   “NextGenerationEU”/PRTR. 
IDM is also supported by the project PID2021-122938NB-I00 funded by the Spanish “Ministerio de Ciencia e Innovación” and FEDER “A way of making Europe", and by the project SA096P20 Junta de Castilla y León.

\bibliographystyle{bibtex/aa} 
\bibliography{biblio}

\begin{thebibliography}{46}
\expandafter\ifx\csname natexlab\endcsname\relax\def\natexlab#1{#1}\fi

\bibitem[{{Arvanitaki} {et~al.}(2010){Arvanitaki}, {Dimopoulos}, {Dubovsky},
  {Kaloper}, \& {March-Russell}}]{Arvanitaki2010}
{Arvanitaki}, A., {Dimopoulos}, S., {Dubovsky}, S., {Kaloper}, N., \&
  {March-Russell}, J. 2010, \prd, 81, 123530

\bibitem[{{Bertone} \& {Hooper}(2018)}]{Bertone2018}
{Bertone}, G. \& {Hooper}, D. 2018, Reviews of Modern Physics, 90, 045002

\bibitem[{{Bonnet et al.}(2004)}]{Bonnet2004}
{Bonnet et al.} 2004, The Messenger, 117, 17

\bibitem[{{Broadhurst} {et~al.}(2020){Broadhurst}, {De Martino}, {Luu},
  {Smoot}, \& {Tye}}]{Broadhurst2020}
{Broadhurst}, T., {De Martino}, I., {Luu}, H.~N., {Smoot}, G.~F., \& {Tye}, S.
  H.~H. 2020, \prd, 101, 083012

\bibitem[{{Chen} {et~al.}(2017){Chen}, {Schive}, \& {Chiueh}}]{Chen2017}
{Chen}, S.-R., {Schive}, H.-Y., \& {Chiueh}, T. 2017, \mnras, 468, 1338

\bibitem[{{Davies} \& {Mocz}(2020)}]{Davies2020}
{Davies}, E.~Y. \& {Mocz}, P. 2020, \mnras, 492, 5721

\bibitem[{{Davoudiasl} \& {Denton}(2019)}]{Davoudiasl2019}
{Davoudiasl}, H. \& {Denton}, P.~B. 2019, \prl, 123, 021102

\bibitem[{{De Martino} {et~al.}(2020){De Martino}, {Broadhurst}, {Henry Tye},
  {Chiueh}, \& {Schive}}]{DeMartino2020PDU}
{De Martino}, I., {Broadhurst}, T., {Henry Tye}, S.~H., {Chiueh}, T., \&
  {Schive}, H.-Y. 2020, Physics of the Dark Universe, 28, 100503

\bibitem[{{De Martino} {et~al.}(2017){De Martino}, {Broadhurst}, {Tye},
  {Chiueh}, {Schive}, \& {Lazkoz}}]{DeMartino2017}
{De Martino}, I., {Broadhurst}, T., {Tye}, S. H.~H., {et~al.} 2017, \prl, 119,
  221103

\bibitem[{{de Martino} {et~al.}(2020){de Martino}, {Chakrabarty}, {Cesare},
  {Gallo}, {Ostorero}, \& {Diaferio}}]{deMartino2020}
{de Martino}, I., {Chakrabarty}, S.~S., {Cesare}, V., {et~al.} 2020, Universe,
  6, 107

\bibitem[{{De Martino} {et~al.}(2021){De Martino}, {della Monica}, \& {De
  Laurentis}}]{deMartino2021}
{De Martino}, I., {della Monica}, R., \& {De Laurentis}, M. 2021, \prd, 104,
  L101502

\bibitem[{{Della Monica} \& {de Martino}(2022)}]{DellaMonica2022}
{Della Monica}, R. \& {de Martino}, I. 2022, \jcap, 2022, 007

\bibitem[{{Della Monica} {et~al.}(2022){Della Monica}, {de Martino}, \& {de
  Laurentis}}]{DellaMonica2022b}
{Della Monica}, R., {de Martino}, I., \& {de Laurentis}, M. 2022, \mnras, 510,
  4757

\bibitem[{{Do et al.}(2019)}]{Do2019}
{Do et al.} 2019, Science, 365, 664

\bibitem[{{Eisenhauer} {et~al.}(2003){Eisenhauer}, {Sch{\"o}del}, {Genzel},
  {Ott}, {Tecza}, {Abuter}, {Eckart}, \& {Alexander}}]{Eisenhauer2003}
{Eisenhauer}, F., {Sch{\"o}del}, R., {Genzel}, R., {et~al.} 2003, \apjl, 597,
  L121

\bibitem[{{Foreman-Mackey} {et~al.}(2013){Foreman-Mackey}, {Hogg}, {Lang}, \&
  {Goodman}}]{emcee}
{Foreman-Mackey}, D., {Hogg}, D.~W., {Lang}, D., \& {Goodman}, J. 2013, \pasp,
  125, 306

\bibitem[{{Genzel} {et~al.}(2010){Genzel}, {Eisenhauer}, \&
  {Gillessen}}]{Genzel2010}
{Genzel}, R., {Eisenhauer}, F., \& {Gillessen}, S. 2010, Reviews of Modern
  Physics, 82, 3121

\bibitem[{{Gillessen} {et~al.}(2009){Gillessen}, {Eisenhauer}, {Trippe},
  {Alexander}, {Genzel}, {Martins}, \& {Ott}}]{Gillessen2009}
{Gillessen}, S., {Eisenhauer}, F., {Trippe}, S., {et~al.} 2009, \apj, 692, 1075

\bibitem[{{Gillessen} {et~al.}(2017){Gillessen}, {Plewa}, {Eisenhauer}, {Sari},
  {Waisberg}, {Habibi}, {Pfuhl}, {George}, {Dexter}, {von Fellenberg}, {Ott},
  \& {Genzel}}]{Gillessen2017}
{Gillessen}, S., {Plewa}, P.~M., {Eisenhauer}, F., {et~al.} 2017, \apj, 837, 30

\bibitem[{{Goodman} \& {Weare}(2010)}]{Goodman2010}
{Goodman}, J. \& {Weare}, J. 2010, Communications in Applied Mathematics and
  Computational Science, 5, 65

\bibitem[{{Gravity Collaboration}(2017)}]{Gravity2017}
{Gravity Collaboration}. 2017, \aap, 602, A94

\bibitem[{{Gravity Collaboration}(2018)}]{gravity2018}
{Gravity Collaboration}. 2018, \aap, 615, L15

\bibitem[{{Gravity Collaboration}(2020)}]{Gravity2020}
{Gravity Collaboration}. 2020, \aap, 636, L5

\bibitem[{{GRAVITY Collaboration}(2021)}]{Gravity2021}
{GRAVITY Collaboration}. 2021, arXiv e-prints, arXiv:2112.07478

\bibitem[{{Grould} {et~al.}(2017){Grould}, {Vincent}, {Paumard}, \&
  {Perrin}}]{Grould2017}
{Grould}, M., {Vincent}, F.~H., {Paumard}, T., \& {Perrin}, G. 2017, \aap, 608,
  A60

\bibitem[{{Hayashi} {et~al.}(2021){Hayashi}, {Ferreira}, \&
  {Chan}}]{Hayashi2021}
{Hayashi}, K., {Ferreira}, E. G.~M., \& {Chan}, H. Y.~J. 2021, \apjl, 912, L3

\bibitem[{{Hlo{\v{z}}ek} {et~al.}(2018){Hlo{\v{z}}ek}, {Marsh}, \&
  {Grin}}]{Hlozek2018}
{Hlo{\v{z}}ek}, R., {Marsh}, D. J.~E., \& {Grin}, D. 2018, \mnras, 476, 3063

\bibitem[{{Hlozek} {et~al.}(2015){Hlozek}, {Grin}, {Marsh}, \&
  {Ferreira}}]{Hlozek2015}
{Hlozek}, R., {Grin}, D., {Marsh}, D. J.~E., \& {Ferreira}, P.~G. 2015, \prd,
  91, 103512

\bibitem[{Hui {et~al.}(2017)Hui, Ostriker, Tremaine, \& Witten}]{Hui2017}
Hui, L., Ostriker, J.~P., Tremaine, S., \& Witten, E. 2017, Phys. Rev. D, 95,
  043541

\bibitem[{{Kawai} {et~al.}(2022){Kawai}, {Oguri}, {Amruth}, {Broadhurst}, \&
  {Lim}}]{Kawai2022}
{Kawai}, H., {Oguri}, M., {Amruth}, A., {Broadhurst}, T., \& {Lim}, J. 2022,
  \apj, 925, 61

\bibitem[{{Kobayashi} {et~al.}(2017){Kobayashi}, {Murgia}, {De Simone},
  {Ir{\v{s}}i{\v{c}}}, \& {Viel}}]{Kobayashi2017}
{Kobayashi}, T., {Murgia}, R., {De Simone}, A., {Ir{\v{s}}i{\v{c}}}, V., \&
  {Viel}, M. 2017, \prd, 96, 123514

\bibitem[{{M. Cautun et al.}(2020)}]{Cautun2020}
{M. Cautun et al.} 2020, \mnras, 494, 4291

\bibitem[{{Mocz} {et~al.}(2017){Mocz}, {Vogelsberger}, {Robles}, {Zavala},
  {Boylan-Kolchin}, {Fialkov}, \& {Hernquist}}]{Mocz2017}
{Mocz}, P., {Vogelsberger}, M., {Robles}, V.~H., {et~al.} 2017, \mnras, 471,
  4559

\bibitem[{{N. K. Porayko et al.}(2018)}]{Porayko2018}
{N. K. Porayko et al.} 2018, \prd, 98, 102002

\bibitem[{{P. Salucci et al.}(2021)}]{Salucci2021}
{P. Salucci et al.} 2021, Frontiers in Physics, 8, 579

\bibitem[{{Pozo} {et~al.}(2021){Pozo}, {Broadhurst}, {de Martino}, {Luu},
  {Smoot}, {Lim}, \& {Neyrinck}}]{Pozo2021}
{Pozo}, A., {Broadhurst}, T., {de Martino}, I., {et~al.} 2021, \mnras, 504,
  2868

\bibitem[{{Rogers} \& {Peiris}(2021)}]{Rogers2021}
{Rogers}, K.~K. \& {Peiris}, H.~V. 2021, \prl, 126, 071302

\bibitem[{{Safarzadeh} \& {Spergel}(2020)}]{Safarzadeh2020}
{Safarzadeh}, M. \& {Spergel}, D.~N. 2020, \apj, 893, 21

\bibitem[{{Schive} {et~al.}(2014{\natexlab{a}}){Schive}, {Chiueh}, \&
  {Broadhurst}}]{Schive2014}
{Schive}, H.-Y., {Chiueh}, T., \& {Broadhurst}, T. 2014{\natexlab{a}}, Nature
  Physics, 10, 496

\bibitem[{{Schive} {et~al.}(2014{\natexlab{b}}){Schive}, {Liao}, {Woo}, {Wong},
  {Chiueh}, {Broadhurst}, \& {Hwang}}]{Schive2014b}
{Schive}, H.-Y., {Liao}, M.-H., {Woo}, T.-P., {et~al.} 2014{\natexlab{b}},
  \prl, 113, 261302

\bibitem[{{Schutz}(2020)}]{Schutz2020}
{Schutz}, K. 2020, \prd, 101, 123026

\bibitem[{{Taff} \& {Szebehely}(1986)}]{Taff1986}
{Taff}, L.~G. \& {Szebehely}, V. 1986, {Celestial Mechanics - a Computational
  Guide for the Practitioner}, ed. Wiley-Interscience, Vol. 319, 630

\bibitem[{{The Event Horizon Telescope
  Collaboration}(2022{\natexlab{a}})}]{EHT2022a}
{The Event Horizon Telescope Collaboration}. 2022{\natexlab{a}}, \apjl, 930,
  L12

\bibitem[{{The Event Horizon Telescope
  Collaboration}(2022{\natexlab{b}})}]{EHT2022f}
{The Event Horizon Telescope Collaboration}. 2022{\natexlab{b}}, \apjl, 930,
  L17

\bibitem[{Vesely(2001)}]{Vesely2001}
Vesely, F.~J. 2001, Computational Physics, 2nd edn. ("New York, NY": Kluwer
  Academic/Plenum)

\bibitem[{{Will}(2008)}]{Will2008}
{Will}, C.~M. 2008, \apjl, 674, L25

\end{thebibliography}

\begin{appendix} 
   \section{Equations of motion}
   \label{app:eqiations of motion}

   We assume that the dynamics of the S2 star around the Galactic Center can be described in the weak field limit of GR. In this regime, the effects on the acceleration of S2 (assumed as a point particle) arising from the central SMBH and those coming from the halo of FDM can be linearly added. In particular, the 1PN order acceleration term experienced by the particle due to the field of a Schwarzschild black hole is given by \citep{Gravity2020, Will2008}
   \begin{equation}
       \vec{a}_{\bullet} = -\frac{GM_\bullet}{r^3}\vec{r}+\frac{GM_\bullet}{c^2r^2}\left[\left(4\frac{GM_\bullet}{r}-v^2\right)\frac{\vec{r}}{r}+4\dot{r}\vec{v}\right]\,,
       \label{eq:PPN_acc_app}
   \end{equation}
   where $M_\bullet$ is the mass of the central SMBH and $\vec{r}$ and $\vec{v}$ are the position and the velocity of the test particle respectively. On the other hand, we consider the acceleration due to the FDM halo to be directed radially (under the assumption that the mass centroid of the dark matter distribution coincides with the SMBH) and thus $\vec{a}_{\rm DM} = a_{\rm DM}^r\vec{r}/r$, where this acceleration is related to the gravitational potential produced by this mass distribution obtained from the density profile via the Poisson equation, namely
   \begin{equation}
       \nabla^2V_s = 4\pi G \rho_s(r)\,,
   \end{equation}
   which we can express in spherical coordinates (assuming that the potential only depends on $r$), yielding to
   \begin{equation}
       \frac{1}{r^2}\frac{\partial}{\partial r}\left(r^2\frac{\partial V_s}{\partial r}\right) = 4 \pi G\frac{\rho_0}{(1+Ar^2)^8}\,.
       \label{eq:acc_pn_app}
   \end{equation}
   where the constants $\rho_0$ and $A$ are related to the mass $m_a$ of the FDM boson, as reported in the main text. This equation can be integrated analytically obtaining directly the radial component of the acceleration experienced by test particles due to the presence of the bosonic dark matter halo made:
   \begin{widetext}
   \begin{align}
       -a^r_{\rm DM}=\frac{dV_s}{dr} =& \frac{4\pi G\rho_0}{r^2}\int \frac{r^2}{(1+Ar^2)^8}dr=\nonumber\\
       =&\frac{\pi G \rho_0[(3465 A^6 r^{12} + 23100 A^5 r^{10} + 65373 A^4 r^8 + 101376 A^3 r^6 + 92323 A^2 r^4 + 48580 A r^2 - 3465)]}{53760 Ar (1+A r^2)^7}+\nonumber\\&+\frac{33\pi G \rho_0 \tan^{-1}(\sqrt{A}r)}{512A^{3/2}r^2}\,.
       \label{eq:acc_dm}
   \end{align}
   \end{widetext}
   The total acceleration term is thus given by:
   \begin{equation}
       \vec{a}=\vec{a}_\bullet+\vec{a}_{\rm DM}.
   \end{equation}
   The orbit of S2, hence, can be obtained by integrating the equations of motion (Newton's second law):
   \begin{equation}
       \ddot{\vec{r}} = \vec{a}.
       \label{eq:eq_motion}
   \end{equation}
   \section{Initial conditions}
   \label{app:initial_conditions}
   
   These equations can be integrated upon assigning initial conditions at a given time $t_0$ for the position and velocity of the star. It can be seen that, due to the spherical symmetry of the gravitational field, if one assigns as initial conditions in spherical coordinates $\theta(t_0) = \pi/2$ and $\dot{\theta}(t_0) = 0$, one obtains $\ddot{\theta} = 0$, identically. This implies that one can always define the reference frame so that the motion of the test particle occurs in the equatorial plane. For the sake of convenience, we start our integration at the last apocenter passage $t_0=t_p-T/2\sim\textrm{J}2010.35$ and fix initial conditions at that time by means of the Keplerian orbital elements of the star. In more detail, the orbital period $T$, the time of pericenter passage $t_P$, the semi-major axis $a$ and the eccentricity $e$ serve the purpose of fixing the in-orbital-plane motion of the star, by identifying a unique Keplerian ellipse on the orbital plane. Due to the 1PN term in the acceleration and to the presence of the extended mass component around the central point-mass, however, the orbit is not a closed ellipse, but experiences an orbital precession. This means that the orbital elements correspond to the ones that identify the ellipse that osculates the true trajectory at a given time. We use the classical relations between the orbital elements and the position and velocity of the star at a given time (see the appendices of \cite{Grould2017} and references therein), to fix the initial position and velocity of S2 on the equatorial plane. Once established the appropriate initial conditions, we have integrated Eqs. \eqref{eq:eq_motion} numerically, by means of first-order Euler symplectic (semi-implicit) scheme \citep{Vesely2001}. This results in a parametric array that describes the motion of S2 in a reference frame centered on the central source of the gravitational field. Since the equations of motion that we integrate contain both the post-Newtonian term in \eqref{eq:PPN_acc_app} and the extended mass contribution \eqref{eq:acc_dm}, all the relativistic effects on the orbital trajectory related to the central compact mass (e.g. its general relativistic pericenter advance) and perturbations introduced by the DM halo are naturally taken into account in our synthetic orbit. However, in order to be able to compare the integrated orbit with the observational data, a projection is required in the reference frame of a distant observer, by means of the following relations \citep{Taff1986}:
   \begin{align}
        x&=\mathcal{B}x_{\rm orb}+\mathcal{G}y_{\rm orb}\,,\\
        y&=\mathcal{A}x_{\rm orb}+\mathcal{F}y_{\rm orb}\,,\\
        z&=\mathcal{C}x_{\rm orb}+\mathcal{H}y_{\rm orb}+D_\bullet\,.\label{eq:z_obs}
   \end{align}
   where $(x_{\rm orb},y_{\rm orb})$ are the coordinates of the star on its orbital plane, while $(x,y)$ and $z$ are the sky-projected position of the star and its distance from the observer, respectively (here, $D_\bullet$ represents the galactocentric distance of the observer, that we leave as a free parameter). The constants $\mathcal{A}$, $\mathcal{B}$, $\mathcal{C}$, $\mathcal{F}$, $\mathcal{G}$ and $\mathcal{H}$ are obtained from the inclination $i$ of the orbit, the angle of the line of nodes $\Omega$ and the argument of the pericenter $\omega$, via \citep{Taff1986}
   \begin{align}
       & \mathcal{A}=\cos\Omega\cos\omega-\sin\Omega \sin\omega \cos i\,,\\
       & \mathcal{B}=\sin\Omega \cos\omega+\cos\Omega \sin\omega \cos i\,,\\
      & \mathcal{C}=-\sin\omega \sin i\,,\\
       & \mathcal{F}=-\cos \Omega \sin\omega-\sin\Omega \cos\omega \cos i\,,\\
       & \mathcal{G}=-\sin\Omega\sin\omega+\cos\Omega \cos\omega \cos i\,,\\
       & \mathcal{H}=-\cos\omega \sin i\,.
   \end{align}
   A similar expression leads from the velocities on the orbital plane $(v_{\rm x, orb}, v_{\rm y, orb})$ to the sky-projected ones $(v_x, v_y)$ and the line-of-sight velocity $v_z$ \citep{Taff1986}:
   \begin{align}
       v_x&= \mathcal{B}v_{\rm x, orb}+\mathcal{G}v_{\rm y, orb}\,,\\
       v_y &= \mathcal{A}v_{\rm x, orb}+\mathcal{F}v_{\rm y, orb}\,,\\
      v_z &= -(\mathcal{C}v_{\rm x, orb}+\mathcal{H}v_{\rm y, orb})\,.\label{eq:line_of_sight_v}
   \end{align}
   Here, we put a minus sign in front of \eqref{eq:line_of_sight_v} as we adopt the convention that the line-of-sight velocity is positive when the celestial body approaches the observer, and negative during the receding phase \citep{Grould2017}. Additionally, in order to properly reconstruct synthetic orbits for S2, we have to take into account the classical Rømer delay experienced by light rays transiting over the orbit, which affects both the sky-projected positions and the line-of-sight velocity (we refer to expression reported in \citet{Grould2017} to quantify such delay). Furthermore, the line-of-sight velocity, observable through spectroscopic observations, is affected by additional redshift contributions coming from special relativistic transverse Doppler effect and the general  relativistic gravitational time dilation (we refer to \citet{deMartino2021} and \citet{DellaMonica2022b}, for a detailed analysis on how these effects can be quantified on our predicted orbits). Finally, higher-order relativistic effects produced by the central mass (e.g. Shapiro time delay or gravitational lensing) are not considered as their contribution to the S2 star astrometric observation is expected to be below the current instrumental sensitivity (see for example \citet{Grould2017} and \cite{DellaMonica2022}) and the extremely diluted (with respect to the central mass) FDM field is expected to give a negligible contribution to such effects.
   
   \section{Mock data for S2}
   \label{app:mock}
   
   In order to constrain our orbital model, we used a mock catalog of observations of the S2 star mirroring the observational limitations of the VLT instruments GRAVITY (for the astrometry) and SINFONI (for the spectroscopic measurements leading to radial velocity estimates). This mock catalog was first presented in \cite{DellaMonica2022} and we report its main features here. Firstly, we consider the motion of S2 around a Schwarzschild BH by integrating numerically the relativistic (geodesic) equations of motion. This is done by considering  the orbital parameters resulting from the analysis by the Gravity Collaboration \citep{Gravity2020} as our fiducial ones. Finally, we assume that the GRAVITY instruments perform observations as follows:
   
   \begin{itemize}
     \item one observation per day in the two weeks centered on the pericenter passage;
     \item one observation every two nights in a month centered on the pericenter passage;
     \item one observation per week in the two months centered on the pericenter;
     \item two observations per year in the rest of the years.
   \end{itemize}
   
    Moreover, we assume that spectroscopic and astrometric measurements are always performed on the same nights and we extend the integration over 2 entire periods of the S2. It is furthermore assumed that both instruments are always with their nominal uncertainty of $\sigma_{A} =$ 10 $\mu$as for the GRAVITY astrometry and $\sigma_{RV} = $10 km/s for the SINFONI radial velocities. For more details we refer to Section 4.1 of \cite{DellaMonica2022}.
   
   \section{Data analysis}
   \label{app:data_analysis}
   
   The methodology that we have employed in order to place constraints on the 9-dimensional parameter space of our model is based on the Bayesian sampling of the posterior probability distribution of such parameters. More specifically, we made use of a Markov Chain Monte Carlo (MCMC) algorithm, implemented in \texttt{emcee} \citep{emcee}. Prior probability distributions for each parameter are assigned (see Table \ref{tab:priors}) and the sampler draws random values within these priors. An orbit for S2, corresponding to the parameter sample extracted from the prior distributions, is computed and compared to our mock catalog by computing the log-likelihood, given by
   \begin{equation}
      \log\mathcal{L} = -\frac{1}{2}\sum_i\biggl[\biggl(\frac{x_{\rm mock}^i-x^i}{\sigma_A}\biggr)^2+
      \biggl(\frac{y_{\rm mock}^i-y^i}{\sigma_A}\biggr)^2+\biggl(\frac{v_{\rm z, mock}^i-v_z^i}{\sigma_{RV}}\biggr)^2\biggr]\,.
   \end{equation}
   Here, ($x_{\rm mock}$, $y_{\rm mock}$, $v_{z, \rm mock}$) are the positions and radial velocities for S2 from our mock catalog, while ($x$, $y$, $v_z$) are the ones from our orbital model. We assess the convergence of the posterior sampling with the estimation of the autocorrelation time of the Markov chains \citep{Goodman2010}. 
   
   \begin{table}
   \setlength{\tabcolsep}{19pt}
   \renewcommand{\arraystretch}{1.5}
   \centering
   \begin{tabular}{|lll|}
   \hline
   Parameter                   & $\mu$      & $\sigma$ \\ \hline
   $D_\bullet$ (kpc)           & 8.2467     & 0.093    \\
   $T$ (yr)                    & 16.0455    & 0.013    \\
   $t_p$ (yr)                  & J2018.37800 & 0.00017   \\
   $a$ (mas)                   & 125.058    & 0.044     \\
   $e$                         & 0.884649   & 0.000079   \\
   $i$  ($^\circ$)             & 134.567    & 0.033     \\
   $\omega$  ($^\circ$)        & 66.263     & 0.030    \\
   $\Omega$  ($^\circ$)        & 228.171    & 0.031    \\ \hline
   & \multicolumn{2}{c|}{Uniform interval} \\              
   $\log_{10}[m_a(eV)]$                  & \multicolumn{2}{c|}{ [$-23,-17$]}                                    \\ \hline
   \end{tabular}
   \caption{The sets of priors used in our analysis. The orbital parameters priors come from  \cite{Gravity2020}. In particular, for each parameter, we have adopted a uniform prior over an interval that is $15$ times larger than the corresponding $1\sigma$ confidence interval reported in the table (corresponding to the range [$\mu-15\sigma$, $\mu+15\sigma$] for each parameter). Finally, for the parameter of interest, $m_a$, we have adopted a uniform prior over the interval $10^{-23}\div10^{-17}$ eV that we have sampled logarithmically.}
   \label{tab:priors}
   \end{table}
   
   \section{Results of the posterior analysis}
   \label{app:results}
   
\begin{figure*}
       \centering
       \includegraphics[width = \textwidth]{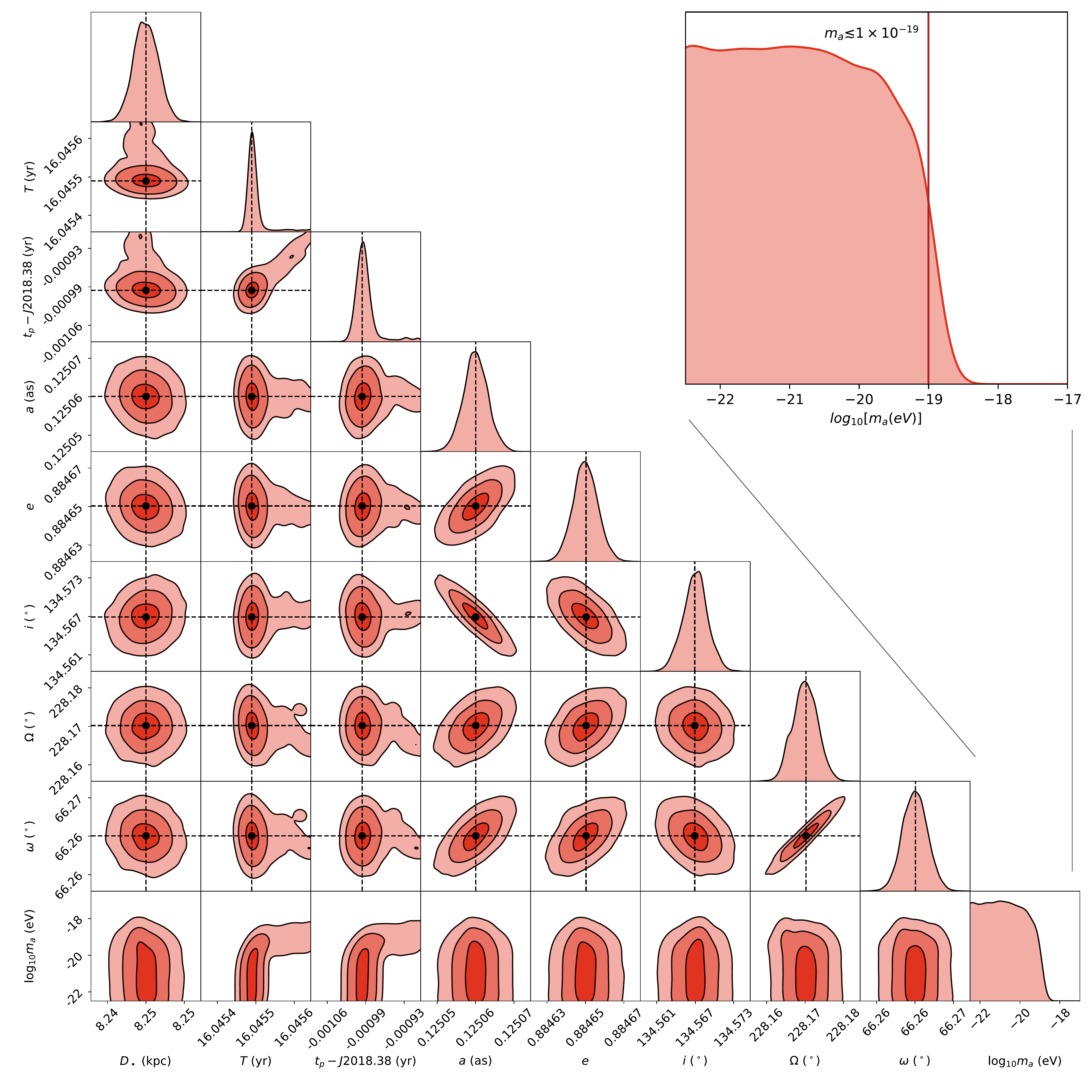}
       \caption{The 9-dimensional posterior distribution resulting from the MCMC analysis. Contours report the 68\%, 95\% and 99.7\% regions on each pair of parameters, while the density plot on top of each column report the single-parameter marginalized distribution. All the parameters of our orbital model are bound and show closed contours, with the exception of $m_a$ for which we can only set an upper limit from our analysis. The black dashed lines and filled circles show the input parameters used to build the mock catalog. The inset reports the marginalized posterior distribution of the parameter $m_a$ in the logarithmic scale. The red vertical line corresponds to the 95\% upper limit of the parameter resulting from our posterior analysis.}
       \label{fig:mcmc_posterior_full}
   \end{figure*}

   The results of our posterior analysis on the 9-dimensional parameter space of our orbital model for S2 are reported in Figure \ref{fig:mcmc_posterior_full} and in Table 1 of the main paper. In particular, the contours plot report the $68-95-99.7\%$ confidence regions for each pair of parameters. The orbital elements of S2 result to be bounded and compatible with the prior values of our mock catalogue, as shown by the black dashed lines and filled circles depicted in Figure \ref{fig:mcmc_posterior_full}. Additionally, the MCMC analysis allows us to place an upper limit on the boson mass which, at 95\% confidence level results to be $m_a\lesssim 1\times10^{-19}$ eV. The marginalized posterior probability density function of the parameter $m_a$ is reported in logarithmic scale in the inset of  Figure \ref{fig:mcmc_posterior_full} for the sake of completeness.
   
\end{appendix}

\end{document}